\documentclass[preprint,showpacs,preprintnumbers,amsmath,amssymb]{revtex4-1}
\usepackage{graphicx}
\usepackage{dcolumn}
\usepackage{bm}
\usepackage{epsfig}
\usepackage{booktabs,topcapt}
\begin{document}


\title{Diffusion anomaly and dynamic transitions in the Bell-Lavis water model
 }

\author{Marcia M. Szortyka\footnote[2]{e-mail - szortyka@gmail.com}}
\affiliation{Departamento de F\'{\i}sica, Universidade Federal de Santa Catarina, 
Caixa Postal 476, 88010-970, Florian\'opolis, SC, Brazil}

\author{Carlos E. Fiore\footnote[1]{e-mail - fiore@fisica.ufpr.br}}
\affiliation{Departamento de F\'isica, Universidade Federal do Paran\'a, 
Caixa Postal 19044, 81531 Curitiba, PR, Brazil}

\author{Vera B. Henriques\footnote[3]{e-mail - vhenriques@if.usp.br}}
\affiliation{Instituto de F\'{\i}sica, Universidade de S\~ao Paulo,
Caixa Postal 66318, 05315970, S\~ao Paulo, SP, Brazil}

\author{Marcia C. Barbosa\footnote[4]{e-mail - marcia.barbosa@ufrgs.br}}
\affiliation{Instituto de F\'isica, Universidade Federal do Rio Grande do
Sul, Caixa Postal 15051, 91501-970, Porto Alegre, RS, Brazil}

\date{\today}
\begin{abstract}
In this paper we investigate  the
dynamic properties  of the minimal Bell-Lavis (BL) water model 
and their relation to the thermodynamic anomalies. The Bell-Lavis
 model is defined on a triangular lattice in
which water molecules are represented by particles with three symmetric bonding
arms interacting through van der Waals and hydrogen bonds.
We have studied the model diffusivity in different regions of the phase diagram
through Monte Carlo simulations. 
Our results show that the model displays a region of anomalous diffusion which
lies inside the region of anomalous density, englobed by the line
of temperatures of maximum density (TMD).
Further, we have found that the diffusivity undergoes a dynamic transition
which may be classified as fragile-to-strong transition at the
critical line only at low pressures. At higher densities, no dynamic transition
is seen on crossing the critical line. Thus evidence from this study is that
relation of dynamic transitions
to criticality may be discarded. 
 \end{abstract}
\maketitle

\section{Introduction}

Water is  the most familiar substance in nature, and nonetheless
a satisfactory understanding of its properties is still
lacking.
Many of its properties are regarded as anomalous, when compared with those of other
substances \cite{URL}. Its most well-known peculiar property
 is probably the density anomaly \cite{An76},
which increases with temperature for a range of pressures.
In addition, different response functions, such as specific heat,
isothermal compressibility and thermal expansion 
coefficient also display peculiar behaviors.

Besides thermodynamic anomalies, water also exhibits
dynamic anomalies, seen in both experiments \cite{An76} and in simulations
 \cite{Ne01}. In usual fluids, diffusivity increases
with decreasing density, since mobility is enhanced in a less dense medium. 
However, in the case of liquid water, a range
of pressures exists for which diffusivity exhibits non monotonic behavior with
density, and both minima and a maxima in the diffusion coefficient may be found. 

It has been proposed a few years ago that these anomalies would be related to the second critical point
between two liquid phases, the low-density liquid (LDL) and the high-density liquid (HDL) phases \cite{Po92}.
This critical point, discovered through computer simulations, might be located in the supercooled
region beyond the line of homogeneous nucleation and is thus unacessible experimentally.
 This hypothesis has been supported by indirect experimental results 
\cite{Mi98}. In
spite of the limit of $235\; K$ below which water cannot be found in the liquid phase without
crystallization, two amorphous phases were observed at much lower temperatures \cite{Mi84}. 
There is some evidences, even if not
definitive, of the presence 
of the   two liquid phases \cite{Mi02,Ma04,Sm99}.

Recently, experimental results in nanoscale hydrophilic pores
show a crossover from fragile to strong diffusivity as temperature
is lowered,in the supercooled region,
at constant pressure \cite{Li04,Li05,Xu05}.
The concept of fragility, introduced by Angell \cite{An97}, classifies
the liquids as strong or fragile, whether the diffusion coefficient 
displays Arrhenius or non-Arrhenius behavior, respectively.
In order to give further support to the 
hypothesis of a critical point at the end
of the coexistence
line between the two liquid phases,  it
was suggested that
this crossover from a fragile to strong regime in water would
signal the presence of criticality. In particular, it was proposed
that the fragile-to-strong transition observed in water is 
associated with crossing the Widom line, the analytic continuation
of the coexistence line.

Is  the fragile-to-strong transition associated to the
presence of criticality? In order
to address this question,  a number of
models which display criticality were
investigated as to the presence of fragile-to-strong transitions
 \cite{Ku08,Sz09,Sz10}. These studies have shown that on crossing
the critical line,  fragile-to-strong, strong-to-strong
or even fragile-to-fragile
transitions could be observed, depending on the specific structure
of the phases separated by the critical line. In the particular
case of the Associated Lattice Gas Model, which presents two critical lines,
two kinds of dynamic transitions are also present. The  critical line separating 
the fluid from the low density liquid phase, at lower pressures, could
be associated to a fragile-to-strong transition, 
whereas the critical line separating the high density
fluid from the high density liquid phase, at higher pressures,
was associated to a strong-to-strong 
dynamic transition \cite{Sz09}. In both cases, the
the dynamic transition is of the same kind along the whole critical line.
Thus a logical question arises: is the type of the dynamic transition linked with
the universality class of the critical line? or does it only
depend on the nature of the phases related to the 
dynamic transition?

In this paper, we test these ideas on a very simple model
that exhibits a single critical line separating two fluids.
If the universality class of the critical line and the class
of the dynamic transition are associated, we would expect
the model to display dynamic transitions of one class only.
We investigate the diffusion properties of the 
 Bell Lavis (BL) water
model \cite{Be70},
the only two dimensional
 ice-like  orientational model known to us
which does not require an energy penalty in order to
 present a density anomaly. It is a triangular lattice gas model in
which water molecules are represented by particles with three symmetric bonding
interacting through van der Waals and hydrogen bonds.
It is probably the simplest orientational  model that
reproduces water like anomalies.
Our study will focus on  three questions: are dynamic
anomalies and dynamic transitions verified in a minimal model?
If present, how are they related to thermodynamic anomalies? Are dynamic
transitions related the criticality?

This paper is organized as follows: in Sec. II the model is described
and its phase diagram is reviewed; in Sec. III the simulation
results
for the model dynamic anomalies and dynamic
transitions are presented; Sec V resumes our conclusions.

\section{The Bell-Lavis model and phase diagram}
\begin{figure}[h!]
\begin{center}
\includegraphics[clip=true,scale=0.3]{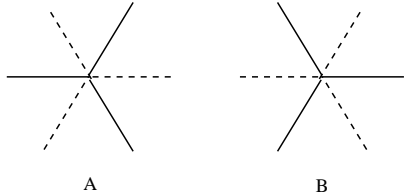}
\caption{Two possible particle orientation configurations. Solid lines are the bonding arms while dashed lines are non-bonding arms.\label{mol}}
\end{center}
\end{figure}

The Bell-Lavis model is a two-dimensional  system in which
 molecules are located on a triangular lattice and 
are represented by two kinds of variable, in order to represent occupational and
orientational states. The occupational variables $\sigma_{\scriptscriptstyle{i}}$ 
assume the value $\sigma_{\scriptscriptstyle{i}}=0$, if the site is empty, and 
$\sigma_{\scriptscriptstyle{i}}=1$, if the site is occupied by a molecule. 
The orientational variables $\tau_i{\scriptscriptstyle{i}}^{ij}$ are introduced to represent
the possibility of bonding between molecules.
Each molecule has six arms, separated by 120$^{\circ}$, three of them inert, with $\tau_{\scriptscriptstyle{i}}^{ij}=0$, 
while the other three are the bonding arms, with 
$\tau_{\scriptscriptstyle{i}}^{ij}=1$.
The two possible orientations $A$ and $B$ for the molecule are illustrated in 
Fig. \ref{mol}.

Two neighbor molecules interact via van der 
Waals and hydrogen bonding.
The model energy is described by the following 
effective Hamiltonian, in the grand-canonical ensemble: 
\begin{equation}
{\cal H}=-\sum_{(i,j)}
\sigma_{i}\sigma_{j}(\epsilon_{hb}\tau_{i}^{ij}\tau_{j}^{ij}+\epsilon_{vdw})
-\mu \sum_{i}\sigma_{i},
\label{hamiltoniana}
\end{equation}
where $\epsilon_{hb}$  and
$\epsilon_{vdw}$ are  the  strength of hydrogen bond (hb) and  
van der Waals (vdW) interaction energies, respectively and 
$\mu$ is the chemical potential.

The phase diagram of this model was investigated
for different values of the bonding strength, with
different approaches: under a mean-field
approach \cite{Be70,La73,Ba08}, 
with renormalization group techniques \cite{Yo79,So80} 
and very recently, through detailed numerical
simulations \cite{Fi09}. In this paper, we restrict our analysis to
two values of the bonding strength parameter $\zeta\equiv \epsilon_{vdw}/\epsilon_{hb}$,
$\zeta=1/10$ and $\zeta=1/4$. These two parameter values
are interesting because in both cases the system exhibits
two liquid phases. However, for  $\zeta=1/10$,
the critical line ends at a tricritical point, while for 
$\zeta=1/4$ it does end at a 
critical end point. 
 
\begin{figure}[h!]
\includegraphics[clip=true,scale=0.4]{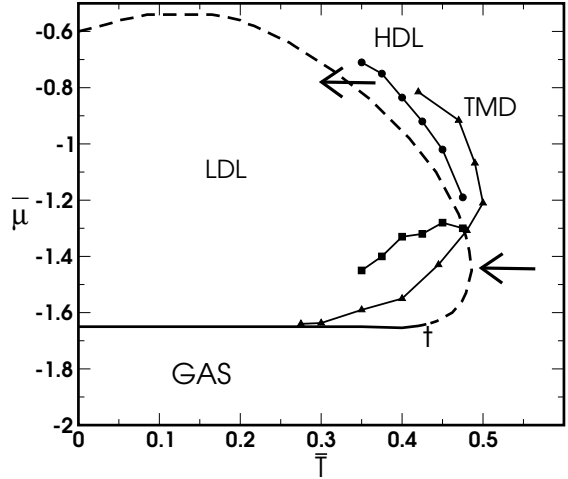}
\caption{Chemical potential $\overline{\mu}$ vs. temperature  
$\overline{T}$ phase diagram for $\zeta=1/10$.                                                    
The solid line is a first order transition line between the gas and the LDL phases. The dashed line 
is a second order transition line between the LDL and the HDL phases. 
 The point $t$ is a tricritical point. Triangles 
are points of density maxima and the continuous line represents the TMD line. 
Circles  and squares are diffusivity
maxima and  minima locci, respectively.}
\label{fig1}
\end{figure}

\begin{figure}[h!]
\includegraphics[clip=true,scale=0.4]{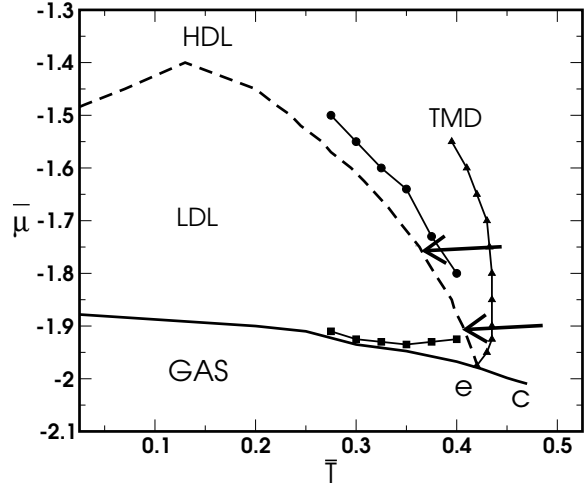}
\caption{Chemical potential $\overline{\mu}$ vs. temperature  
$\overline{T}$ phase diagram for $\zeta=1/4$.                                    
 The solid line is a first order transition line between the gas 
and the LDL phases. The dashed line 
is a second order transition line between the LDL and the HDL phases. 
The points $e$ and $c$ correspond to end-critical point and 
 critical point, respectively. Triangles 
are points of density maxima and the continuous line represents the TMD line. Circles and squares are diffusivity
maxima and minima  locci, respectively. }
\label{fig2}
\end{figure}

The chemical potential ${\bar \mu}$
versus temperature ${\bar T}$ model phase diagram 
 is shown in Figs.~\ref{fig1} and ~\ref{fig2},
 for $\zeta=1/10$ and $\zeta=1/4$, respectively.  Reduced 
units for temperature and chemical potential are defined as
\begin{equation}
\overline{T} = \frac{T}{\epsilon_{hb}} 
\qquad {\rm and } \qquad \overline{\mu}=\frac{\mu}{\epsilon_{hb}}.
\label{reduced}
\end{equation}


\begin{figure}[h!]
\setlength{\unitlength}{1.0cm}                                    
\includegraphics[clip=true,scale=0.7]{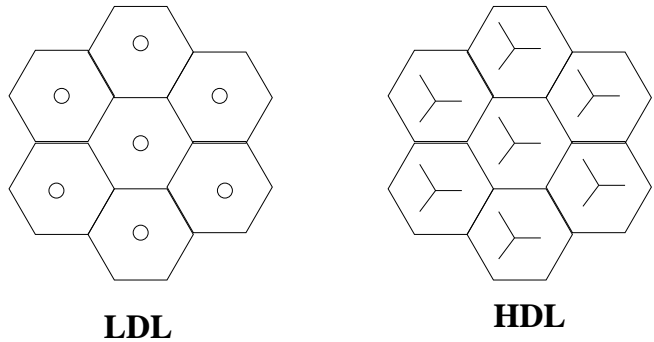}
\caption{Typical bond configurations for the 
LDL and the HDL phases at $\overline{T}=0$.}
\label{phases}
\end{figure}

For both $\zeta=1/10$ and $\zeta=1/4$, the
system displays three different phases. For low chemical potential,
the system is constrained in the gas phase, with density $\rho\approx 0$. 
For intermediate values of the chemical potential, the system is 
in the low density liquid phase (LDL). For high 
chemical potentials, the system exhibits a high density liquid 
phase (HDL). The LDL and HDL phases are separated by a critical line,
which has been identified as an order-disorder transition  \cite{Fi09}.  
Typical configurations for the zero temperature LDL and HDL configurations
 are illustrated  in Fig.~\ref{phases}. 

The phase transition between gas and LDL phases is first-order for
both values of $\zeta$ \cite{Fi09}. For this 
transition, the order-parameter is associated to 
density $\rho=n/V$, where $n$ is 
the number of occupied sites 
while $V=L^2$ is 
the number of sites. At zero temperature, the two phases coexist, with 
$\rho \approx 0$ for the gas, 
and $\rho \approx 2/3$ for the LD liquid.
For higher bonding strength, $\zeta=1/10$, the coexistence line 
ends at a  tricritical point, whereas for lower bonding strength, $\zeta=1/4$, 
it ends at a critical point. 
The HDL-LDL critical line ends at the coexistence line,
thus yielding coexistence between the HDL and the gas
phases. The phase transition between LDL-HDL phases is second-order for both
value of  $\zeta$ and has been associated to an orientational order-disorder 
 transition \cite{Fi09}. 

In  both cases of smaller and larger bonding strengths, the system 
displays a region of anomalous thermodynamic behavior.
For  $\zeta=1/4$, the line of temperatures of maximum  density (TMD)
is located inside the HDL phase. For  $\zeta=1/10$, it crosses the LDL
phase, for lower pressures, and migrates to the HDL phase, for high
 pressures.

\section{Diffusivity and dynamic transitions}

We have studied diffusivity for the Bell-Lavis model over its phase diagram
through Monte Carlo simulations.

The numerical algorithm for studying mobility is described as follows:
(i) the system is equilibrated with fixed chemical potential (or density) 
and fixed temperature; (ii) an occupied site $i$ and it's neighbor $j$
are chosen randomly; (iii) if the neighbor site $j$ is empty, the molecule 
moves to the empty site and the difference between the final and the initial 
energy $\Delta E$ is computed; (iv) if $\Delta E<0$, the movement is 
accepted, otherwise 
the movement is accepted with a probability $\exp(\Delta E /\bar k_{B} T)$. 
A Monte Carlo step is defined through the number of trials of movement 
for every particle.  
After repeating this algorithm $nt$ times, where $n$ is the number of 
molecules in the lattice, the diffusion coefficient is evaluated 
according to Einstein's equation
\begin{equation}                        
\overline {D}=\lim_{t \rightarrow \infty} \frac{\langle \Delta
  r(t)^{2}\rangle}{4t},        
\label{eq1}
\end{equation}
where $\langle \Delta r(t)^{2}\rangle=\langle (r(t)-r(0))^{2} \rangle$
is the  mean square displacement per particle and time is measured in Monte Carlo steps.

Our data have been obtained for lattice 
size $L=18$ under periodic boundary conditions.

\subsection{Diffusion anomaly}
In normal liquids, the diffusion coefficient grows as the density decreases. 
However, in  anomalous liquids,  the diffusivity decreases from a maximum at 
$\rho_{\scriptscriptstyle{D_{max}}}$ to a minimum 
at $\rho_{\scriptscriptstyle{D_{min}}}$, as the density is decreased. 
For densities outside this region the diffusion behaves as described
above, i. e, as a normal liquid.

\begin{figure}[h!]
\setlength{\unitlength}{1.0cm}                                    
\includegraphics[clip=true,scale=0.35]{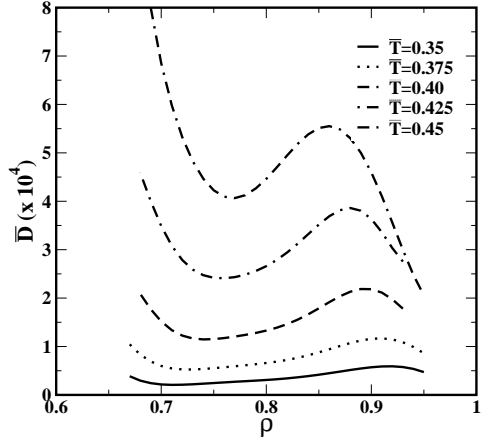}
\caption{Reduced diffusion $\overline{D}$ versus density $\rho$ 
for several temperatures
$\overline{T}$ ranging from 0.350 to 0.450  for $\zeta=1/10$.}
\label{fig3}
\end{figure}

\begin{figure}[h!]
\setlength{\unitlength}{1.0cm}
\includegraphics[clip=true,scale=0.35]{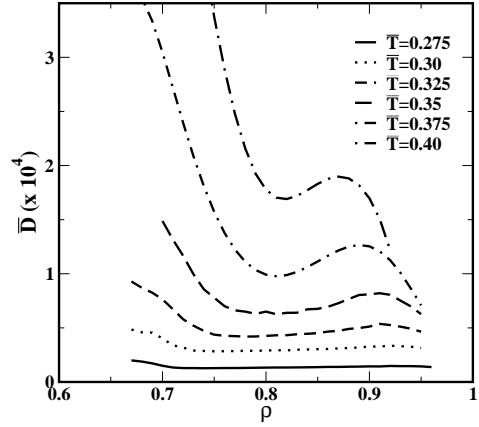}
\caption{Reduced diffusion $\overline{D}$ versus density $\rho$ for
  several temperatures $\overline{T}$ ranging from 0.275 to 0.400  for 
$\zeta=1/4$.}
\label{fig4}
\end{figure}

In order to investigate the existence of this anomaly for the BL model,
the diffusion coefficient was computed as a function of density, for 
fixed temperatures, for both $\zeta=1/10$ and $\zeta=1/4$. The results 
are shown in Figs.~\ref{fig3} and  \ref{fig4}. 

For $\zeta=1/10$, the diffusion coefficient 
exhibits a maximum in the region
 $0.82<\rho_{\scriptscriptstyle{D_{max}}}<0.92$ 
and  a minimum for $0.71<\rho_{\scriptscriptstyle{D_{min}}}<0.78$,
and temperatures between 0.35 and 0.45. 
For densities lower than $\rho_{\scriptscriptstyle{D_{min}}}$ diffusivity 
behaves 
normally, increasing as the 
 density decreases. A similar behavior is  verified for $\zeta=1/4$, with the
diffusivity maximum located in the interval
 $0.87<\rho_{\scriptscriptstyle{D_{max}}}<0.94$, and the minimum diffusivity
in the range $0.74<\rho_{\scriptscriptstyle{D_{min}}}<0.82$, for temperature 
 interval  ranging from 0.275 and 0.400.

\begin{figure}[h!]
\setlength{\unitlength}{1.0cm}
\includegraphics[clip=true,scale=0.35]{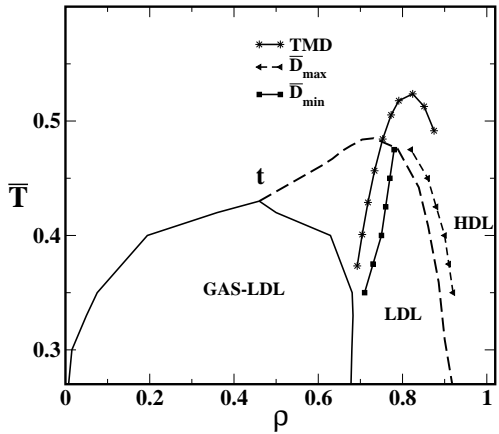}
\caption{Phase diagram $\overline{T}$ versus $\rho$  for
$\zeta=1/10$. Continuous and dashed lines correspond to the coexistence
and critical lines, respectively, which meet at the tricritical
point $t$. Stars correspond to the 
density maxima, whereas  squares and triangles denote the locci of diffusivity
maxima and minima, respectively.}
\label{fig5}
\end{figure}

\begin{figure}[h!]
\setlength{\unitlength}{1.0cm}
\includegraphics[clip=true,scale=0.35]{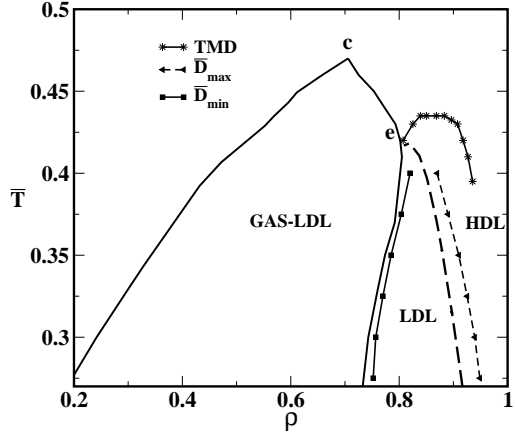}
\caption{Phase diagram $\overline{T}$ versus $\rho$  for       
$\zeta=1/4$. Circles and dashed lines correspond to the coexistence      
line between gas-LDL phase and critical lines, 
respectively that meet at the end                     
point $e$, giving rise to the LDL-HDL phase coexistence. Stars correspond to 
the anomaly density and squares 
denote the regions where the diffusion is  maximum and minimum.}
\label{fig6}
\end{figure}

The locci of maxima and minima in diffusivity 
define a region of diffusion anomaly in the phase diagram, which 
are illustrated in Figs.~\ref{fig5} and 
\ref{fig6}, for $\zeta=1/10$ and $\zeta=1/4$ in the  space 
of ${\o T}$ versus $\rho$ 
and in  Figs. \ref{fig1} and \ref{fig2} 
in the space of chemical potential ${\o \mu}$. 
As can be seen,  the maximum
in diffusivity is located just above the critical line in the HDL phase, whereas 
the minimum in diffusivity is within the low density liquid, close to the
gas-liquid coexistence line.
Thus the diffusion anomalous region lies across the LDL-HDL
critical line. Since the LDL phase is characterized by  bonds ordering, 
this explains the loss of particle mobility, as the LDL phase is approached
from the HDL bond-disordered phase. Note that loss in mobility
initiates in the bond-disordered phase close to the critical line,  
may being related
to large fluctuations in bonding density. On the other hand, inside the LDL
phase, as density is further lowered, mobility again increases, in spite
of bond order, probably due to the large increase of vacant sites
vacant sites, as the 2/3 density of the fully translationally ordered
phase is approached.

A point to note further is that the anomalous diffusion region is enveloped by
the border of the region of density anomaly. This is different
 from the behavior presented by liquid water, but is common to
other lattice models
 \cite{Sz07,Sz09,Sz10}. 
A possible reason for this discrepancy is the fact
that bonding is more rigid in the lattice model, thus reducing the
mobility of particles as compared to continuous models, in which
rotations allow for slightly distorted bonds.

\subsection{Dynamic transitions}

In order to verify the existence of dynamic transitions and its possibility
of relation to  the criticality, 
 diffusivity was computed as a function 
of temperature, for fixed chemical potentials.
The present 
analysis has been carried out in different regions of the phase diagram.
Results for two different chemical potentials have been 
presented for both  stronger and
weaker bonding strength cases ($\zeta=1/10$ and $\zeta=1/4$).

\begin{figure}[h!]
\setlength{\unitlength}{1.0cm}
\includegraphics[clip=true,scale=0.4]{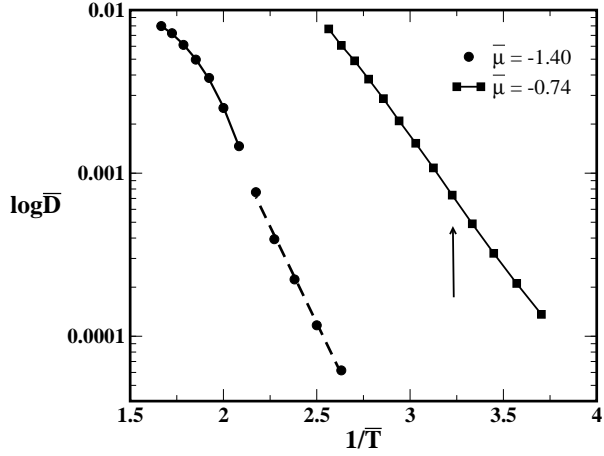}
\caption{For $\zeta = 1/10$ diffusivity undergoes a dynamic transition for chemical potential 
$\bar \mu = -1.40$ as the critical line is crossed. Surprisingly, 
as the same critical line is crossed at higher chemical potential, 
$\bar \mu = -0.74$, diffusivity is not affected. Critical temperatures
are indicated by arrows. 
\label{dz10}}
\end{figure}

For $\zeta=1/10$, the behavior of the diffusion coefficient $\bar D$
with temperature was
analyzed for chemical potentials $\bar \mu = -1.40$ and 
$\bar \mu = -0.74$. The two values of $\bar\mu$ chosen are indicated
by arrows in the phase diagram of Fig.~\ref{fig1}.
 Fig.~\ref{dz10} shows $\bar D$ vs. $1/\bar T$
for the two cases.  For the lower chemical 
potential, the diffusion coefficient undergoes a dynamic 
transition at
the critical line: at high temperatures, diffusivity follows
non-Arrhenius polynomial behavior, given generally by
$\overline{D}=A_{0}+A_{1}\overline{T}+A_{2}\overline{T}^{2}+A_{3}\overline{T}^{3}$, 
which characterizes the system as a fragile liquid; in the low temperature 
region, diffusivity follows an Arrhenius law given by
$\overline{D}=B_{0}\exp(-B_{1}/\overline{T})$, thus characterizing the system as
a strong liquid. The coefficients A$_i$ and B$_i$ are fitting parameters, which
are not investigated in this study.
Surprisingly, for the higher chemical potential 
$\bar \mu = -0.74$ the dynamic crossover at the critical line 
is no longer detected. 
 At this chemical potential, the critical line is 
crossed at a temperature $\bar T \approx 0.31$ (1/$\bar T \approx 3.22$) and, 
as can be seen in Fig.~\ref{dz10}, the system is insensible to the presence of 
the critical line.

%

For $\zeta = 1/4$ the behavior of 
the diffusivity
was also analyzed for two different chemical potentials,
$\bar \mu = -1.90$ and $\bar \mu = -1.75$, as indicated by arrows 
in the phase diagram of Fig.~\ref{fig1}.  
The assumptions for the weaker hydrogen bonding case is
 similar to that of the stronger bonding one, as shown in Fig. \ref{dz4}.
A dynamic transition is seen only for the lower 
chemical potential $\bar \mu = -1.90$, whereas for the 
higher chemical potential,$\bar \mu = -1.75$,
diffusivity is no longer affected by the presence of the critical line,
 showing once more that despite the system presenting a thermodynamic
phase transition, no significant 
structural changing has occurred in the latter case, 
implying therefore in the absence of  dynamic transition.

\begin{figure}[h!]
\setlength{\unitlength}{1.0cm}
\includegraphics[clip=true,scale=0.4]{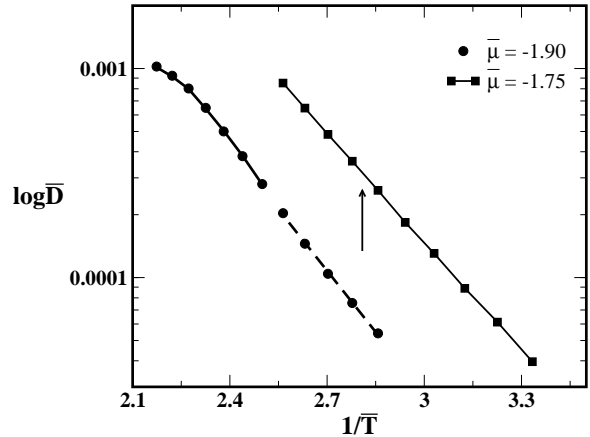}
\caption{For $\zeta = 1/4$ diffusivity undergoes a dynamic transition for chemical potential 
$\bar \mu = -1.90$ at the critical line. Similar to the 
$\zeta = 1/10$ system, diffusivity is no longer affected by crossing the critical line 
at a higher chemical potential $\bar \mu = -1.75$. 
Critical temperatures
are indicated by arrows. \label{dz4}}
\label{dz4}
\end{figure}

\section{Conclusions}

In this paper we have addressed the question of the relation 
between critical lines and dynamic
transitions.  In order to highlight the answer to this question we        
we  have investigated  the dynamic behavior of the 
 Bell-Lavis water model. This model has been considered
 because it presents a relative single 
critical line that separates two fluid phases of different structure. Our study
focuses on the diffusion anomaly and dynamic transitions, and on their 
relation to criticality. 
In the first analysis, 
we have found that, in similarly to other two length scales
models, the BL model presents a diffusion anomalous region 
inside the region of density anomalies \cite{Sz09,Sz07}.

Second, we looked for
dynamic transitions by
 analyzing the behavior of diffusivity with temperature
across the critical line, at 
fixed chemical potentials. Our results showed that two
different regimes may be found: if the critical line is
crossed at low chemical potential, near the minimum
in diffusivity, a fragile-to-strong 
transition is observed;   for higher 
chemical potentials, near the diffusivity maximum, no dynamic transition is seen.  
Thus, different dynamic behavior is seen upon crossing distinct segments
of the same critical line. Our 
explanation for this result is that the structural difference 
on both sides of the critical line, in the region of
higher chemical potential, is 
not enough to provoke a change in diffusivity. 

In summary, our results 
indicate that dynamic transition and criticality are
not directly associated. Instead, the fragile-to-strong transition
(and possibly strong-to-strong or even a fragile-to-fragile
transitions)
is the result of an expressive change in the structure
of the liquid and of polymorphism \cite{An07}.

\section*{ACKNOWLEDGMENTS}

We thank for financial support the Brazilian science agencies
CNPq and Capes. This work is partially supported by CNPq, INCT-FCx.

\bibliographystyle{aip}
\bibliography{Biblioteca}

\end{document}